\documentclass
[twocolumn,showpacs,preprintnumbers,amsmath,amssymb,prc]{revtex4}
\usepackage{graphicx}
\usepackage{dcolumn}
\usepackage{bm}

\begin{document}
\title{Effect of hyperons on phase coexistence in strange matter}

\author{P. Das, S. Mallik and G. Chaudhuri}

\affiliation{Theoretical Nuclear Physics Group, Variable Energy Cyclotron Centre, 1/AF Bidhan Nagar, Kolkata 700064, India}

\begin{abstract}
The  study of liquid gas phase transition in fragmentation of nuclei in heavy ion collisions has been extended to the strangeness sector using the statistical model for multifragmentation. Helmholtz's free energy, specific heat and few other thermodynamic observables have been analyzed in order to examine the occurrence of phase transition in the strange matter. The bimodal behaviour of the largest cluster formed in  fragmentation also strongly indicates  coexistence of both the phases. The presence of hyperons strengthens the signals and also shifts the transition temperature to lower values.
\end{abstract}


\maketitle
{\bf {\it Introduction:-}}The study of liquid-gas phase transition is an important area of research in the regime of fragmentation of nuclei in heavy ion collisions \cite{Siemens}. This is a well studied subject both both theoretically and experimentally \cite{Dasgupta_Phase_transition,Borderie2,Gross_phase_transition,Mallik10} which has a pivotal role  in understanding the physics of nuclei as well as nuclear matter\cite{Siemens,Chomaz}. Extensive research has been done in the past few decades in this topic but this has been mainly confined to normal nuclei or non strange matter. Theoretical and experimental study of nuclear systems with finite strangeness i.e. hypernuclei in heavy ion reactions is a hot topic in the nuclear and particle physics community \cite{Kermann,Dover,Wakai,Chrien,Schaffner1,Schaffner2,Feliciello}. Recently in the past few years the study of phase coexistence has been extended to the domain of strange matter in nuclei \cite{Bielich1,Oertel,Wang,Mallik12,Torres} as well as in (proto) neutron star matter \cite{Francesca_PNS,Bielich2}.\\
\indent
Multifragmentation of normal nuclei \cite{Das,Bondorf1} serves as a important tool to study first order phase transition in nuclei as well as the phase diagram in nuclear matter. In fact the fragmentation of nuclei at a certain temperature range is identified with the well studied phenomenon of liquid gas phase transition. Depending on their size, the produced fragments are  believed to constitute the finite system counterpart of coexisting phase of gas and liquid. This study has been extended to the strangeness sector through the fragmentation of hypernuclei\cite{Mallik12,Dasgupta_hypernuclei1,Dasgupta_hypernuclei2,Botvina1,Botvina2,Botvina3} where the composition of the produced hyper(strange) fragments can serve as a tool for studying liquid-gas phase transition. The role of the strange particles(hyperons)affecting this decomposition is an interesting subject of study in the recent day activities related to fragmentation of hypernuclei. It has already been emphasized that the hyperons have a tendency to get attached to the heavier fragments after break-up\cite{Mallik12}. The U-shaped distribution of the fragments observed in the fragmentation of non-strange nuclei is also exhibited in the fragmentation of hypernuclei and it is this feature which has been exploited in order to study the first order phase transition of the hyperfragments.\\
\indent
Recent study of the liquid-gas phase transition of hypernuclei has been performed \cite{Mallik12} using the three component canonical thermodynamical model( neutron, proton and hyperon) \cite{Dasgupta_hypernuclei1,Dasgupta_hypernuclei2}. The mass distribution of the fragments which  was the main observable of focus in that study pointed out to signatures of phase transition. The mass distribution was analyzed at different temperatures and also for different strangeness content. Variation of the average mass of the fragments as a function of the strangeness content was studied in detail. The main motivation of this work is to continue with this topic of phase transition in hypernuclei exploring the variation of thermodynamic potentials and the size of the largest cluster. The variation of Helmholtz's free energy and its derivatives  is analyzed in order to comment on the nature of the phase transition of the hypermatter. The probability distribution of the largest cluster is analyzed in order to examine its bimodal behaviour at the transition temperature where the peaks in liquid and the gas phase are expected to be equal in height. The effect of strangeness on the transition temperature is also studied in order to demonstrate the effect of hyperons on the phase transition, the variation of specific heat as well as size of the largest cluster.  The influence of the long range Coulomb interaction on the liquid gas phase transition is a well studied subject and in this work  the same study  is being continued including the hyperons.\\

{\bf {\it Theoretical formalism:-}}  In this section we describe briefly the three component canonical thermodynamical model. Assume that the system with
$A_0$ baryons, $Z_0$ protons and $H_0$ hyperons at temperature $T$, has expanded to a higher than normal volume and thermodynamical(statistical) equilibrium is reached at this freeze-out condition.  In a canonical model, the partitioning into different composites is done such that all partitions have the correct $A_0, Z_0$, $H_0$.  Details of the implementation of the model can be found elsewhere \cite{Mallik12,Dasgupta_hypernuclei1}; here we give the essentials necessary to follow the present work. The canonical partition function is given by
\begin{equation}
Q_{A_0,Z_0,H_0}=\sum\prod\frac{(\omega_{a,z,h})^{n_{a,z,h}}}{n_{a,z,h}!}
\end{equation}
Here the product is over all fragments of one break up channel and sum is over all possible channels of break-up satisfying $A_{0}=\sum a\times n_{a,z,h}$, $Z_{0}=\sum z\times n_{a,z,h}$ and $H_{0}=\sum h\times n_{a,z,h}$; $\omega_{a,z,h}$ is the partition function of one composite with $a$ baryons $z$ protons and $h$ hyperons whereas $n_{a,z,h}$ is the number of this composite in the given channel. The partition function $Q_{A_{0},Z_{0},H_{0}}$ is calculated using the recursion relation \cite{Das,Chase}
\begin{equation}
Q_{A_0,Z_0,H_0}=\frac{1}{A_0}\sum_{a,z,h}a\omega_{a,z,h}Q_{A_0-a,Z_0-z,H_0-h}
\end{equation}
\begin{figure}[b]
\begin{center}
\includegraphics[width=6cm,keepaspectratio=true,clip]{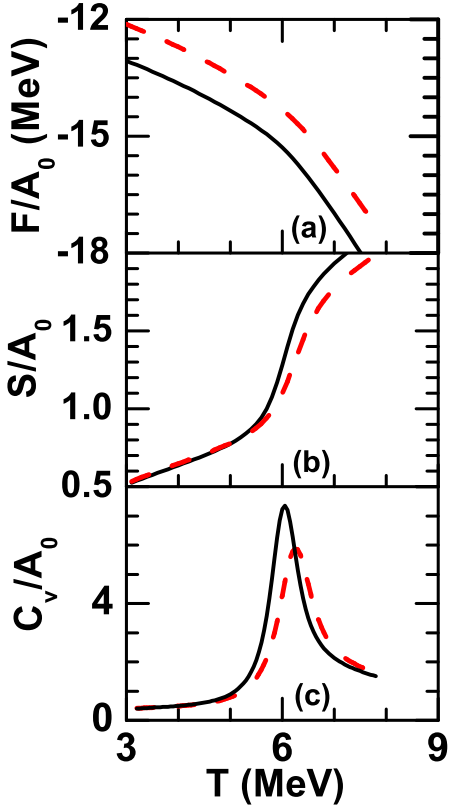}
\caption{Variation of Helmholtz's free energy per nucleon (upper panel), entropy per nucleon (middle panel) and specific heat per nucleon (bottom panel) with temperature for two fragmenting systems having the same $A_0=128$, $Z_0=50$ but different $H_0=8$ (black solid lines) and $H_0=0$ (red dashed lines)}
\label{fig1}
\end{center}
\end{figure}
From Eq. (1) and (2), the average number of composites is given by \cite{Das}
\begin{equation}
\langle n_{a,z,h}\rangle=\frac{\omega_{a,z,h}Q_{A_0-a,Z_0-z,H_0-h}}{Q_{A_0,Z_0,H_0}}
\end{equation}
The partition function of a composite having $a$ baryons, $z$ protons and $h$ hyperons is a product of two parts: one is due to
the the translational motion and the other is the intrinsic partition function of the composite:
\begin{equation}
\omega_{a,z,h}=\frac{V}{h^3}(2\pi T)^{3/2}\{(a-h)m_n+hm_h\}^{3/2}\times z_{a,z,h}(int)
\end{equation}
Here $m_n$ and $m_h$ are masses of nucleon (we use 938 MeV) and hyperon (we use 1116 MeV for $\Lambda$ hyperon) respectively  and $V$ is the volume available for translational motion. Note that $V$ will be less than $V_{f}$, the volume to which the system has expanded at break up \cite{Majumder}. Since hyperfragments are generally studied from projectile like fragments \cite{Das,Bondorf1,Mallik3,Mallik101}, hence we have considered $V_f = 3V_0$.  We use $V=V_{f}-V_{0}$ , where $V_{0}$ is the normal volume of hypernucleus with $A_0$ baryons, $Z_0$ protons and $H_0$ hyperons. The intrinsic partition function arises due to internal motion of the baryons of a fragment which can be calculated from the well known thermodynamic identity $Z_{int}=\exp^{-F/T}$, where $F=E(T)-TS(T)=E_0(T)+E_{ex}(T)-TS(T)$ is the Helmholtz Free energy (as equilibrium is considered at constant volume). The neutron, proton and $\Lambda$ particles are taken as the fundamental blocks therefore $z_{int}(1,0,0)$=$z_{int}(1,1,0)$=$z_{int}(1,0,1)$=1. In order to construct $z_{int}(a,z,h)$ for the different fragments, experimental binding energies are used for the lower mass region ($1<a\le5$) both for $h=$0 and $h>$0 (wherever available) and for $a>$5, liquid-drop formula \cite{Botvina3} is used for obtaining ground state energy and Fermi-gas model is applied for studying excitation energy $E_{ex}(T)$ and entropy $S(T)$. The liquid-drop formula \cite{Botvina3} is given by
\begin{eqnarray}
E_0(T)&=-&16a+\sigma(T)a^{2/3}+0.72kz^2/(a^{1/3}) \nonumber\\
&&+25(a-h-2z)^2/(a-h)-10.68h \nonumber\\
&&+21.27h/(a^{1/3})
\end{eqnarray}
\begin{figure}[b]
\begin{center}
\includegraphics[width=6.4cm,keepaspectratio=true,clip]{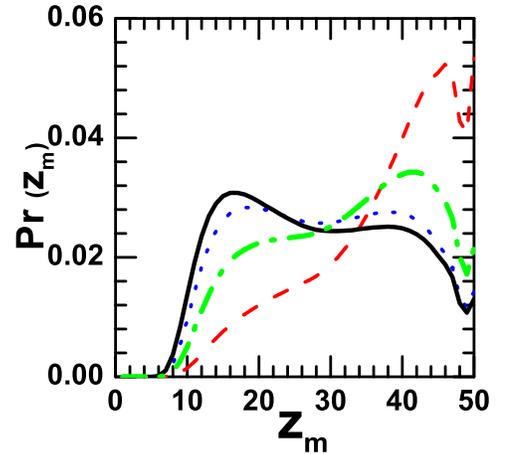}
\caption{Largest cluster probability distribution for three different fragmenting systems having same $A_0=128$, $Z_0=50$ but different $H_0=8$ (black solid line), $H_0=4$ (blue dotted line), $H_0=2$ (green dash dotted line) and $H_0=0$ (red dashed line). Calculations are done at constant temperature $T=$6.065 MeV.}
\label{fig2}
\end{center}
\end{figure}
where $\sigma(T)$ is the surface tension which is given by $\sigma(T)=\sigma_{0}\{(T_{c}^2-T^2)/(T_{c}^2+T^2)\}^{5/4}$ with $\sigma_{0}=18.0$
MeV and $T_{c}=18.0$ MeV and k is the correction factor in Coulomb energy which incorporates the effect of its long-range behavior
by Wigner-Seitz approximation as in Ref. \cite{Bondorf1}.\\
\indent
It is necessary to specify which nuclei are included in computing $Q_{A_{0},Z_{0},H_0}$ [Eq. (2)]. For $1<a\le8$, the we have considered the same set of nuclei as mentioned in Ref. \cite{Dasgupta_hypernuclei2} and for $a>8$ we include a ridge along the line of stability. The liquid-drop formula gives neutron and proton drip lines for different strangeness and the results shown here include all nuclei within the boundaries. Another useful parametrization in liquid drop formula for hypernuclei was proposed
by Samanta et. al. \cite{Samanta}. A comparative study of these two formula in the case of hyperfragmentation was described
in Ref. \cite{Botvina3} and finally the one used here was chosen because it produces results closer to the experimental data.\\
\begin{figure}[!h]
\includegraphics[width=6.4cm,keepaspectratio=true,clip]{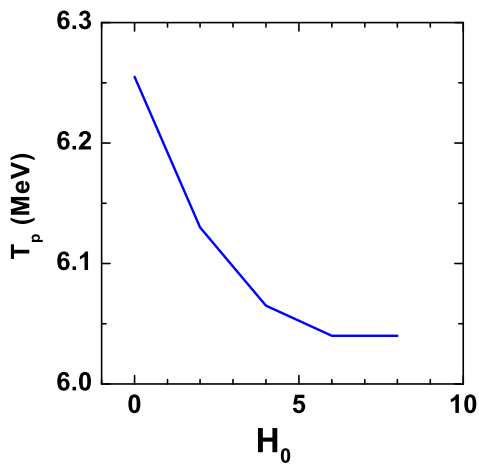}
\caption{Variation of transition temperature ($T_p$) with the total strangeness content of the fragmenting system.}
\label{fig3}
\end{figure}
\indent
{\bf {\it Results and Discussions:-}} Fig. 1.a shows the variation of Helmholtz's free energy ($F=-TlnQ_{A_0,Z_0,H_0}$) with temperature for a system with baryon number $128$, charge $50$. The Coulomb interaction has been switched off in order to better reveal the signatures of phase coexistense  in a finite system. The calculation is done both for normal ($H=0$) nuclei as well as the strange ones with 8 hyperons. There is not much qualitative difference between the two plots and the thermodynamic potential displays a continuous trend. The next plot (Fig 1.b) shows the variation of the first derivative of the free energy with respect to  temperature i.e. entropy $\big{(}S=-{(\frac{\partial F}{\partial T})}_V\big{)}$ which exhibits a discontinuity in its variation with temperature. The variation is more pronounced for the strange system as is evident from the figure. This indicates that the presence of hyperons in the system strengthens the transition process thereby amplifying the signals. There is sharp rise in the derivative in the temperature range of 5.5 to 6 MeV after which it again slows down. Fig 1.c displays the behaviour of the 2nd derivative of the free energy with respect to temperature i.e. specific heat $\big{(}C_v=T{(\frac{\partial S}{\partial T})}_V\big{)}$ for both normal and the strange system which shows a peak as is expected for systems undergoing phase transition. The peak is sharper in the case of strange system as is expected since indications of phase coexistence is more pronounced for multi-hyperon system as compared to the non strange ones. The temperature where the peak appears is different in both cases , the transition temperature being less for the strange system. The reason behind this is probably due to the fact that more hyperons attached to a nucleus is equivalent to the case of a  nucleus carrying more excitation energy and hence it disintegrates faster compared to that of a less strange system.\\
\indent
\begin{figure}
\includegraphics[width=6.4cm,keepaspectratio=true,clip]{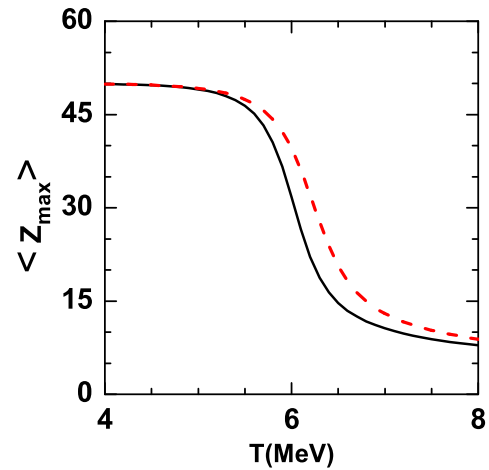}
\caption{Variation of average charge of largest cluster ($Z_{max}$) with temperature ($T$) for two fragmenting systems having the same $A_0=128$, $Z_0=50$ but different $H_0=8$ (black solid line) and $H_0=0$ (red dashed line).}
\label{fig4}
\end{figure}
\begin{figure}[b]
\includegraphics[width=6.4cm,keepaspectratio=true,clip]{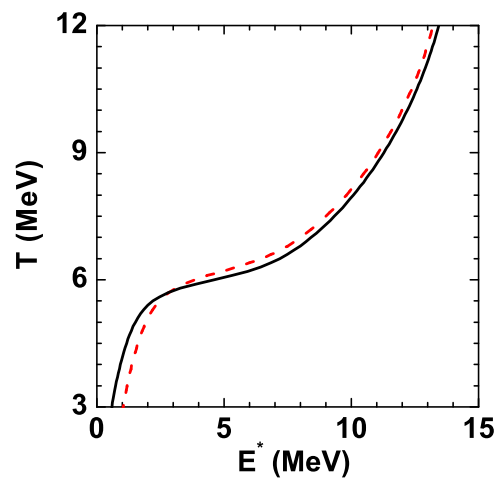}
\caption{Variation of temperature ($T$) with excitation energy ($E^*$) for two fragmenting systems having the same $A_0=128$, $Z_0=50$ but different $H_0=8$ (black solid line) and $H_0=0$ (red dashed line).}
\label{fig5}
\end{figure}
The largest cluster size acts as a very good order parameter for phase transition in nuclear multifragmentation. Fig. 2 displays the probability distribution of the largest cluster size at T=6.065 MeV for systems with varying amount of strangeness content. Here the probability that $z_{m}$ is the largest cluster is calculated from the relation
\begin{equation}
Pr(z_{m})=\frac{Q_{A_0,Z_0,H_0}(z_{m})-Q_{A_0,Z_0,H_0}(z_{m}-1)}{Q_{A_0,Z_0,H_0}(Z_0)}
\end{equation}
\begin{figure}[b]
\includegraphics[width=6.4cm,keepaspectratio=true,clip]{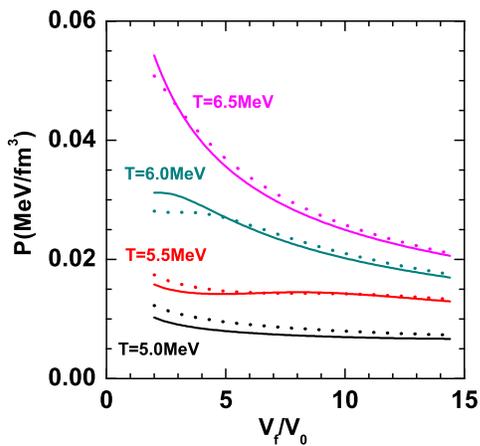}
\caption{Variation of pressure with volume for two fragmenting systems having the same $A_0=128$, $Z_0=50$ but different $H_0=8$ (solid lines) and $H_0=0$ (dashed lines) at four different temperatures $T=$5.0, 5.5, 6.0 and 6.5 MeV.}
\label{fig6}
\end{figure}
where $Q_{A_0,Z_0,H_0}(z_{m})$ can be constructed such that all values of $\omega_{a,z,h}$ are set at $0$ when $z\ge z_{m}$ \cite{Chaudhuri_largest_cluster}. The probability distribution is expected to display a bimodal behaviour in and around the transition temperature with two peaks of equal height characterizing the liquid and the gas phase coexisting at the same temperature and this is attributed to phase coexistence or first order phase transition. The transition temperature where the peaks are of equal height varies with the strangeness content and is less for the more strange system as the system with more hyperons breaks down more easily. The bimodal behaviour \cite{Gulminelli1,Chaudhuri_bimodality,Mallik14,Bonnet_bimodality} of the size of the largest cluster probability distribution further establishes the occurrence of first order phase transition in the system.\\
\indent
Fig. 3 displays the variation of the transition temperature with the total strangeness content of the system and it further confirms the conclusion already obtained from the previous figure that the phase transition temperature is less for the more strange system indicating that the strangeness aids in disintegration of the system.\\
\begin{figure}[b]
\includegraphics[width=5.7cm,keepaspectratio=true,clip]{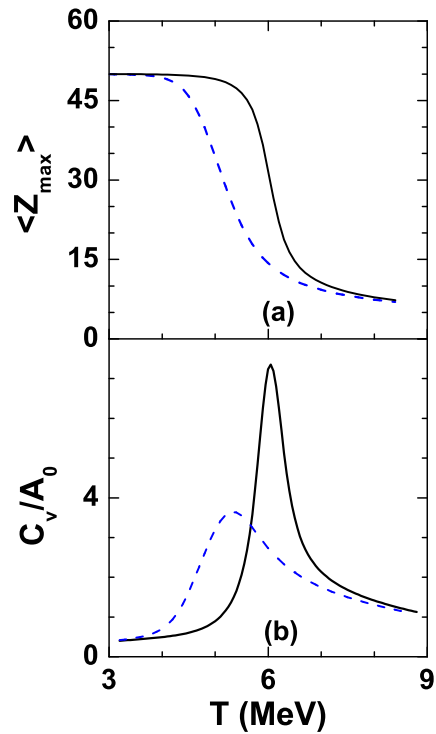}
\caption{Variation of average charge of largest cluster ($Z_{max}$) (upper panel) and specific heat per nucleon (lower panel) with temperature ($T$) by switching on (blue dashed lines) and switching off (black solid lines) the Coulomb interaction. All the calculations are done for the fragmentation of a hypernucleus having $A_0=128$ baryons, $Z_0=50$ protons and $H_0=8$ hyperons.}
\label{fig7}
\end{figure}
\indent
Fig. 4 displays the variation of average charge of the largest cluster $\big{[} \langle Z_{max} \rangle=\sum_{Z_m=1}^{Z_m=Z_0}Z_m Pr(Z_m)\big{]}$ with temperature for both normal nuclei as well  as nuclei with 8 hyperons. There is  a steep change in the average size near the transition temperature  which is again indicative of change from liquid to gas phase. The steepness is more pronounced for the strange system as compared to the normal system.\\
\indent
Fig. 5 shows the caloric curve \cite{Pochodzalla_phase_transition} i.e. the variation of temperature with excitation energy for both normal and strange nuclei switching off the  Coulomb interaction. For a given temperature the excitation energy of the fragmenting system is calculated from the relation
\begin{equation}
E^{*}_{A_0,Z_0,H_0}=T^2\frac{1}{Q_{A_0,Z_0,H_0}}\bigg\{ \frac{\partial Q_{A_0,Z_0,H_0}}{\partial T}\bigg\}-E_{A_0,Z_0,H_0}(T=0)
\end{equation}
\indent
Initially the temperature rises steeply with the excitation, then it slows down during the phase transition process and again starts increasing rapidly. For a system in the thermodynamic limit, temperature is expected to remain constant with excitation energy in the transition region. The nuclei being much smaller in size, signatures are suppressed and one observes a remarkable slowing down in the rate of change of temperature instead of it remaining constant as expected in an infinite system.\\
\indent
Fig 6 shows another important observable related to first order phase transition which are isotherms in the pressure volume plane. The figure displays the variation of pressure with volume for four different temperatures both for normal nuclei as well those with 8 hyperons. Initially the pressure decreases with volume after which it remains  more or less constant irrespective of the change in volume. This is another strong signature of liquid-gas coexistence or first order phase transition. It can also be observed from this figure that for a particular temperature, the strange system disintegrates at a smaller volume as compared to the normal one.\\
\indent
In the next part of our work we attempt to demonstrate the effect of the long range coulomb interaction on phase transition signatures. This topic is well studied in the case of normal nuclei \cite{Qian,Mallik13} but here it is extended to the nuclei with strangeness. One observable on which the effect is being examined is the average size of the largest cluster. The system where Coulomb interaction is considered shows a much slower change in size of $\langle Z_{max} \rangle$ as compared to the case where Coulomb is switched off where the change is much more steep. Hence one can conclude that the effect of Coulomb on signatures of phase transition is similar in both strange  and ordinary nuclei. The next observable which is being considered for testing this is variation of specific heat with temperature. Similar effect is being observed which implies a much broader peak for system with Coulomb as compared to the case without Coulomb where the peak is much narrower. This again confirms that presence of long range force suppresses the indication of phase transition on different relevant observables.\\
{\bf {\it Summary:-}} The study of phase coexistence in normal matter has been extended to the strangeness sector using the three component canonical thermodynamical model. The analysis of the thermodynamical obsevables like free energy, entropy, specific heat clearly points out to the occurrence of phase transition within a certain temperature interval which in turn depends strongly on the strangeness content of the system. The largest cluster size which acts as an order parameter for liquid gas phase transition exhibits a bimodal behaviour in its probability distribution at a certain transition temperature which agrees with that obtained from the thermodynamic variables. More strange is the system, less is the transition temperature for conversion to the gas-like phase. The effect of the long-range Coulomb interaction on the phase transition of starnge matter has also been studied using the specific heat and the size of largest cluster as the relevant observables.\\
{\bf {\it Acknowledgement:-}}
The authors  gratefully acknowledge important discussions with Prof. Francesca Gulminelli, University of Caen, France.

\end{document}